\documentclass[epj]{svjour}

\usepackage{graphicx}
\usepackage{fancyhdr}
\usepackage{amsmath, amssymb}
\def\makeheadbox{{
 }}
\def\mytitle{My title} 
\def\myauthors{My name}  
\def\mytype{My type of session}
\def\mysession{My session}

\def\mytitle{Cosmology of Gravitino LSP Scenario \\
with Right-Handed Sneutrino NLSP} 
\def\myauthors{Koji Ishiwata, Shigeki Matsumoto, and Takeo Moroi}  
\def\mytype{Contributed Talk}    
\def\mysession{Cosmology and Astrophysics}

\begin{document}
\title{Cosmology of Gravitino LSP Scenario \\
with Right-Handed Sneutrino NLSP}

\author{Koji Ishiwata\inst{1},
 Shigeki Matsumoto \inst{1,2},
 \and
 Takeo Moroi \inst{1}
}                     
%
%
\institute{ Department of Physics, Tohoku University, Sendai 980-8578,
Japan \and Tohoku University International Advanced Research and
Education Organization, Institute for International Advanced
Interdisciplinary Research, Sendai, Miyagi 980-8578, Japan}
%
\date{}
\abstract{ We consider supersymmetric model with right-handed
(s)neutrinos where the neutrino masses are purely Dirac-type.  We
discuss cosmology based on such a scenario, paying particular attention
to the case that the gravitino is the lightest superparticles (LSP)
while the right-handed sneutrino is the next-LSP.  It will be shown that
the cosmological constraints on the gravitino-LSP scenario (in
particular, those from the big-bang nucleosynthesis) are drastically
relaxed in such a case.  We will also consider the implication of such
scenario to the structure formation.
\PACS{ 
     {12.60.Jv}{Supersymmetric models} \and 
     {95.35.+d}{Dark matter} \and
     {98.80.Cq}{Particle-theory and field-theory models of the early
     Universe} \and 
     {14.60.Pq}{Neutrino mass and mixing}
     } 
} 
\maketitle
%

\section{Introduction}

Existence of dark matter (DM) in our universe, which is strongly
supported by a lot of recent cosmological observations
\cite{WMAP1,Seljak:2004xh,Spergel:2006hy}, requires physics beyond the
standard model (SM).
Many possibilities of dark matter have been discussed in various
frameworks of particle physics models so far
\cite{ReviewDM}. Importantly, properties of the dark matter particle
depend strongly on the particle physics model we consider.

In the framework of supersymmetry (SUSY), probably most popular
candidate of dark matter is thermally produced lightest neutralino which
is usually assumed to be the lightest superparticle (LSP).  
However, if we try to build a supersymmetric model which
accommodates with all theoretical and experimental requirements, we
expect that there exist new exotic would-be-DM particles which are not
superpartners of the SM particles, for example, gravitino $\psi_{\mu}$
and right-handed sneutrino $\tilde{\nu}_R$, which are superpartners of
graviton and right-handed neutrino, respectively.  If the neutrino
masses are Dirac-type, in particular, $\tilde{\nu}_R$s are expected to
be as light as superpartners of SM particles in the framework of the
gravity-mediated SUSY breaking.  Existence of these exotic
superparticles may significantly change the phenomenology of dark matter
in supersymmetric models.

In this paper, we consider the supersymmetric model in which the
neutrino masses are Dirac-type and discuss cosmological implications of
such a scenario.  It has already been pointed out that the the
$\tilde{\nu}_R$-DM scenario can be realized in
Ref.\cite{AsakaIshiwataMoroi}.  Here, we consider another case
where the gravitino is the LSP and one of the right-handed sneutrinos is
the next-to-the-lightest superparticle (NLSP)\cite{Ishiwata:2007bt}. 
If the gravitino is the
LSP, it may be a viable candidate of dark matter also in the case
without the right-handed sneutrinos
\cite{Moroi:1993mb,FengRajaramanTakayama,Ellis:2003dn}. In such a case,
however, stringent constraints on the scenario are obtained from the
study of the gravitino production at the time of the reheating after
inflation and also from the study of the big-bang nucleosynthesis (BBN)
reactions. With the $\tilde{\nu}_R$-NLSP, we reconsider cosmological
constraints on the $\psi_{\mu}$-LSP scenario. We pay particular
attention to the BBN constraints and also to the constraints from the
structure formation of the universe. We will see that the BBN
constraints are significantly relaxed if there exists the
$\tilde{\nu}_R$-NLSP.


\section{Model Framework}

In this section, we discuss the model in which gravitino is the LSP
while right-handed sneutrino is the NLSP.  We assume that neutrino
masses are purely Dirac-type, and the superpotential of the model is
written as
\begin{eqnarray}
 W
 =
 W_{\rm MSSM} 
 +
 y_{\nu} \hat{L} \hat{H}_u \hat{\nu}^c_R,
\end{eqnarray}
where $W_{\rm MSSM}$ is the superpotential of the minimal supersymmetric
standard model (MSSM), $\hat{L} = (\hat{\nu}_L, \hat{e}_L)$ and
$\hat{H}_u=(\hat{H}^+_u, \hat{H}^0_u)$ are left-handed lepton doublet
and up-type Higgs doublet, respectively. In this article, ``hat" is used
for superfields, while ``tilde" is for superpartners. Generation indices
are omitted for simplicity.

In this model, neutrinos acquire their masses only through Yukawa
interactions as $m_{\nu} = y_{\nu} \langle H^0_{u} \rangle = y_{\nu} v
\sin{\beta}$, where $v$ is the vacuum expectation value (VEV) of the
standard model Higgs field ($v \simeq$ 174 GeV) and $\tan{\beta} =
\langle H^0_u \rangle/\langle H^0_d \rangle$. Thus, the neutrino
Yukawa coupling is determined by the neutrino mass through the
equation:
 $y_{\nu} \sin{\beta}
 =
 3.0 \times 10^{-13}
 \times 
 (
   m^2_{\nu} / 2.8 \times 10^{-3} ~{\rm eV}^2 
 )^{1/2}$.
Mass squared differences among neutrinos have already been determined
accurately at neutrino oscillation experiments \cite{K2K,KamLAND}.  
In this article, we assume that the spectrum of neutrino masses is
hierarchical, hence the largest neutrino Yukawa coupling is of the order
of $10^{-13}$.  We use $y_{\nu}=3.0 \times 10^{-13}$ for the numerical
analysis in this article.  With right-handed (s)neutrinos, new soft SUSY
breaking terms are introduced in addition to the usual terms of the
MSSM. Those are right-handed sneutrino mass terms and tri-linear
coupling terms called $A_\nu$-terms.  Breaking terms relevant to our
analysis are
\begin{eqnarray}
 {\cal L}_{\rm SOFT} 
 &=&
 - M^2_{\tilde{L}}     \tilde{L}^{\dagger} \tilde{L} 
 - M^2_{\tilde{\nu}_R} \tilde{\nu}^{*}_R   \tilde{\nu}_R
\nonumber \\
 && + ( A_{\nu} \tilde{L} H_u \tilde{\nu}^c_R + {\rm h.c} ),
\end{eqnarray}
where all breaking parameters, $M_{\tilde{L}}$, $M_{\tilde{\nu}_R}$
and $A_{\nu}$, are defined at the electroweak (EW) scale. We
parameterize $A_{\nu}$ by using the dimensionless constant $a_\nu$ as
  $A_{\nu} = a_\nu y_{\nu} M_{\tilde{L}}$.
We adopt gravity-mediated SUSY breaking scenario and, in such a case,
$a_\nu$ is expected to be $O(1)$.  Though the $A_{\nu}$-term induces the
left-right mixing in the sneutrino mass matrix, the mixing is safely
neglected in the calculation of mass eigenvalues due to the smallness
of neutrino Yukawa coupling constants. Thus, the masses of sneutrinos
are simply given by 
 $m^2_{\tilde{\nu}_L}
 =
 M^2_{\tilde{L}} + \frac{1}{2} \cos (2 \beta) m^2_Z,
 m^2_{\tilde{\nu}_R}
 =
 M^2_{\tilde{\nu}_R}$, 
where $m_Z \simeq$ 91 GeV is the Z boson mass.  In the following
discussion, we assume that all the right-handed sneutrinos are
degenerate in mass for simplicity.

In this article, we consider the $\psi_{\mu}$-LSP scenario with
$\tilde{\nu}_R$-NLSP.  In such a case, the next-to-next-LSP (NNLSP)
plays an important role in the thermal history of the universe.
However, there are many possibilities of the NNLSP, depending on the
detail of SUSY breaking scenario.  Thus, we concentrate on the case that
the NNLSP is the lightest neutralino, whose composition is
Bino $\tilde{B}$. This situation is easily obtained if we consider
the so-called constrained-MSSM type scenario. It is not difficult to
extend our discussion to the scenario with other NNLSP candidate.

\section{Constraints from BBN}

It is well known that models with the $\psi_{\mu}$-LSP usually receive
stringent constraints from the BBN scenario.  In these models, LSP in
the MSSM sector (which we call MSSM-LSP) is long-lived but decays to
gravitino with hadrons, which may spoil the success of BBN scenario.
The BBN constraints give the upper bound on $Y_{X} E_{\rm vis}$ as a
function of $\tau_{X}$, where $X$ stands for a long-lived but unstable
particle, $Y_{X}\equiv [n_{X}/s]_{t\ll \tau_X}$ (with $n_X$ and $s$
being the number density of $X$ and the entropy density of the universe,
respectively), $E_{\rm vis}$ is the mean energy of visible particles
emitted in the $X$ decay, and $\tau_{X}$ is the lifetime of the particle
$X$ \cite{KawKohMor}. We use the upper bound on $Y_X E_{\rm vis}$
obtained in the studies.
In order to evaluate BBN constraints quantatively, we calculate 
$B_{\rm had} Y_X E_{\rm vis}$ as a function of $\tau_{\tilde{B}}$, then 
search for allowed paramter region.

First, we will see the decay of the MSSM-LSP, which is $\tilde{B}$-like
neutralino, in order to calculate $B_{\rm had} E_{\rm vis}$ and
$\tau_{\tilde{B}}$.  Main decay modes are the following two-body decays:
$\tilde{B} \rightarrow \tilde{\nu}_R \bar{\nu}_L$, $\tilde{B}
\rightarrow \psi_{\mu} \gamma$, and $\tilde{B} \rightarrow \psi_{\mu}
Z$.  By the use of decay widths of these processes in
Ref.\cite{Feng:2004mt}.  
Importanly, the decay mode
$\tilde{B}\rightarrow\tilde{\nu}_R\bar{\nu}_L$ competes with the mode
$\tilde{B} \rightarrow \psi_{\mu} \gamma$ or it even dominates in total
decay in wide parameter region, especially when the gravitino mass
$m_{3/2}$ is larger than 0.1 GeV when $m_{\tilde{\nu}_R}=100~{\rm GeV}$
and $a_{\nu}=1$.  We have also checked that the lifetime of $\tilde{B}$
is $10^{2-3}$ seconds in that parameter region.

Without the $\tilde{\nu}_R$-NLSP, the decay mode $\tilde{B} \rightarrow
\psi_{\mu} \gamma / Z$ dominates in total decay, and many visible
particles are emitted through photon or $Z$ boson, which may spoil the
success of the BBN scenario.  As a result, the gravitino mass is
strictly constrained as $m_{3/2} \lesssim 0.1$ GeV for $\tau_{\tilde{B}}
\lesssim 1$ second \cite{Feng:2004mt}. In our scenario with the
$\tilde{\nu}_R$-NLSP, however, less hadrons are emitted, though the
Bino-like neutralino is long-lived  \footnote{Left-handed neutrinos
injected by the decay might possibly change the abundance of ${}^4$He
\cite{Kanzaki}. However, we have checked that the BBN constraints on the
neutrino injection are much less stringent than those on three- or
four-body decays as shown in the following discussion.}.  Therefore,
constraints from BBN is expected to be relaxed in the $m_{3/2} \gtrsim
0.1$ GeV region.

As we saw that no hadrons are emitted in two-body decays, in order to
calculate $B_{\rm had}E_{\rm vis}$, we consider three- or four-body
decays: $\tilde{B} \rightarrow \psi_{\mu} f \bar{f}$, $\tilde{B}
\rightarrow \tilde{\nu}_R e^+_L f \bar{f}^{\prime}$, and $\tilde{B}
\rightarrow \tilde{\nu}_R \bar{\nu}_L f \bar{f}$, where $f$ and
$\bar{f}$ denote fermion and anti-fermion, respectively.  Although
branching ratios of these processes are much smaller than 1, they have
impacts on the BBN scenario.

Lasty, we determine $Y_{\tilde{B}}$ by the use of following formula for
(would-be) density parameter of $\tilde{B}$ \cite{Feng:2004mt}:
 $\Omega_{\tilde{B}} h^2 
 =
 C_{\rm model} \times 0.1
 \left[
  m_{\tilde{B}} / 100~{\rm GeV} 
 \right]^2$, 
where $m_{\tilde{B}}$ is Bino mass and the additional parameter $C_{\rm
model}$ is introduced to take the model dependence into account in our
analysis: $C_{\rm model} \sim 1$ for the neutralino in the bulk region,
$C_{\rm model} \sim 0.1$ for that in the co-annihilation or funnel
region, and $ C_{\rm model} \sim 10$ for the pure Bino case without
co-annihilation. 


\begin{figure*}
 \begin{center}
  \includegraphics[scale=0.45]{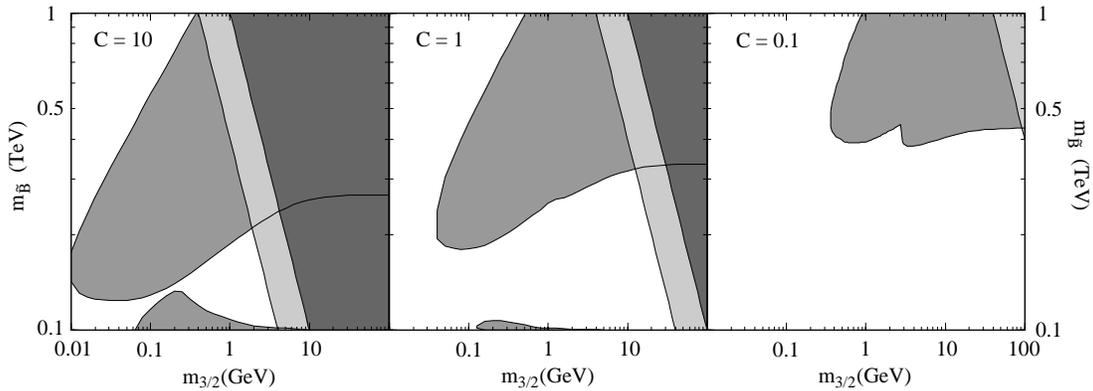}
  \caption{\small Constraints from the BBN on the ($m_{3/2}$,
    $m_{\tilde{B}}$) plane: Parameters are chosen to be
    $m_{\tilde{\nu}_R} = 100$ GeV, $a_{\nu} = 1$, and
    $m_{\tilde{\nu}_L} = 1.5 m_{\tilde{B}}$. We set $C_{\rm model} =
    $10, 1, and 0.1 in the left, middle, and right figures,
    respectively. The middle shaded regions are ruled out by the BBN
    scenario, while dark and light shaded regions are excluded by the
    WMAP measurement and the structure formation of the universe,
    respectively.}
 \label{fig:MgvsMbinoV2-4}
 \end{center}
\end{figure*}

Our numerical results are shown in Fig.\ref{fig:MgvsMbinoV2-4}, where
the BBN constraints are depicted on the ($m_{3/2}$, $m_{\tilde{B}}$)
plane. We take $C_{\rm model} = 10, 1, 0.1$ in the left, middle, and
right figures, respectively. Other parameters are set as
$m_{\tilde{\nu}_R} = 100$ GeV, $a_{\nu} = 1$, and $m_{\tilde{\nu}_L} =
1.5 m_{\tilde{B}}$.  Shaded regions are ruled out by the BBN
scenario. As shown in these figures, the constraints are drastically
relaxed compared to those in models without the $\tilde{\nu}_R$-NLSP.

As shown in the figures, new allowed region appears; for example, for
$C_{\rm model}=1$, 0.1 GeV $\lesssim m_{3/2} \lesssim$ 40 GeV. In that
region, the BBN constraints give the upper bound on
$m_{\tilde{B}}$. Since the decay mode $\tilde{B} \rightarrow \psi_{\mu}
q \bar{q}$ is subdominant in this region, this bound comes mainly from
four-body decays, $\tilde{B} \rightarrow \tilde{\nu}_R e^+_L q
\bar{q}^{\prime}$ and $\tilde{B} \rightarrow \tilde{\nu}_R \bar{\nu}_L q
\bar{q}$. This fact can be understood intuitively: $B_{\rm had}$ and
$E_{\rm vis}$, are enhanced when $m_{\tilde{B}}$ is large. On the
contrary, in the 0.01 GeV $\lesssim m_{3/2} \lesssim$ 0.1 GeV region,
Bino-like neutralino decays mainly into the gravitino through the
$\tilde{B} \rightarrow \psi_{\mu} \gamma$ process with the lifetime
$\tau_{\tilde{B}} \lesssim 1$ second. Since the decay occurs before the
BBN starts, it does not affect the BBN scenario. This situation also
holds in the usual $\psi_{\mu}$-LSP scenario without the
$\tilde{\nu}_R$-NLSP, and the same allowed region can be seen in
Ref.\cite{Feng:2004mt}. In the case of $C_{\rm model} = 10 (0.1)$ in the
left (right) figure, the constraints from the BBN scenario is more
(less) stringent than the $C_{\rm model} = 1$ case. As a result, the
upper bound on $m_{\tilde{B}}$ becomes smaller (larger). We also find
that the region $m_{\tilde{B}} \simeq$ 100 GeV with 0.1 GeV $\lesssim
m_{3/2} \lesssim$ 1 GeV is excluded in the left and middle
figures: this is because the process $\tilde{B} \rightarrow
\tilde{\nu}_R \bar{\nu}_L$ is kinematically suppressed and branching
ratio of the process $\tilde{B} \rightarrow \psi_{\mu} Z$ is relatively
enhanced.

In addition to the BBN constraints, we also depict other cosmological
bounds in Fig.\ref{fig:MgvsMbinoV2-4}: the gravitino abundance
originating in  $\tilde{B}$ must not exceed the value observed
in the WMAP, $\Omega_{\rm DM} h^2 \simeq 0.105$
\cite{Spergel:2006hy},  
which is shown as a dark shaded region in
Fig.\ref{fig:MgvsMbinoV2-4}. This constraint gives the upper bound on
$m_{3/2}$. Another constraint, $\Omega^{\rm dec}_{3/2} <
0.4\Omega_{\rm DM}$, is also depicted as a light shaded region, which
comes from the structure formation of the universe as discussed in the
next section.

\section{Constraints from Structure Formation}

As shown in the previous section, larger value of $m_{3/2}$ is allowed
compared to the case without $\tilde{\nu}_R$-NLSP.  In the newly allowed
parameter region, the MSSM-LSP decays mainly into $\tilde{\nu}_R$, and
$\tilde{\nu}_R$ decays into the gravitino. Since the gravitino is
produced with large velocity dissipation at the late universe, it
behaves as a warm dark matter, and as a result, may affect the structure
formation of the universe. In this section, we consider the constraints
from the structure formation.


$\tilde{\nu}_R$ decays to $\psi_{\mu}$ mainly through two-body process,
$\tilde{\nu}_R \rightarrow \psi_{\mu} \nu_R$.  We find that the lifetime
of $\tilde{\nu}_R$ $\tau_{\tilde{\nu}_R}$ is $10^2$-$10^8$ seconds for
$m_{3/2} = 0.1$-100 GeV with $m_{\tilde{\nu}_R} = 100$ GeV. With the use
of $\tau_{\tilde{\nu}_R}$, we calculate free-streaming length
$\lambda_{\rm FS} \equiv \int^{t_{\rm EQ}}_{\tau_{\tilde{\nu}_R}} dt
v(t)/a(t)$, where $v(t)$ is the velocity of the $\psi_{\mu}$, $a(t)$ is
the cosmic scale factor, and $t_{\rm EQ}$ is the time of the
matter-radiation equality.  As a result, we found $\lambda_{\rm FS}
\simeq 6$ Mpc when $m_{\tilde{\nu}_R}=100$ GeV irrespective of the
$m_{3/2}$.  Thus, it indicates that the component of the dark matter
(i.e., gravitino) from $\tilde{\nu}_R$ decay acts as a warm dark matter
(WDM). In addition to $\tilde{\nu}_R$ decay, gravitinos are also
produced by the thermal scattering at the reheating epoch after
inflation. The abundance of the $\psi_{\mu}$ from the scattering process
is determined by the reheating temperature and the $m_{3/2}$
\cite{Bolz}. Since the $\psi_{\mu}$ from the scattering is
non-relativistic at the time of the structure formation, it acts as a
cold dark matter (CDM). Thus, we have to consider the
constraints from the structure formation of the universe on the WDM+CDM
scenario.



Constraints from the structure formation on the WDM+CDM scenario are
studied in recent works \cite{Viel:2005qj,Kaplinghat}. According to these
studies, it turns out that the matter power spectrum has a step-like
decrease around the free-streaming scale of the WDM component, $k \sim 2
\pi /\lambda_{\rm FS}$. This fact can be understood intuitively, because
only the power spectrum of the WDM component dumps at the scale. 
On the other hand, the power spectrum is estimated from the
observations of the cosmic microwave background
\cite{WMAP1,Spergel:2006hy}, the red shift surveys of galaxies
\cite{Seljak:2004xh}, and so on.
Even though the power spectrum has been experimentally determined
accurately, an ambiguity still
remains when $k^{-1}$ is around 1 Mpc. Therefore, it is not clear
whether the step-like decrease exists or not, if it is small enough. In
this article, we put the conservative constraints on our model: the
power spectrum, to be more precise, the magnitude of the step-like
decrease, should be within the range of the 95\% confidence level of the
observational data \cite{WMAP1}.  This condition gives the upper bound
on the portion of the WDM component.

In our model, the energy density of the dark matter is composed of two
components, $\rho_{\rm DM} = \rho^{\rm dec}_{3/2} + \rho^{\rm
th}_{3/2}$, where $\rho^{\rm dec}_{3/2}$ and $\rho^{\rm th}_{3/2}$ are
the energy densities of gravitino produced by the decay and by the thermal
scattering processes, respectively. Introducing the fraction of WDM
component $f$, we rewrite $\rho_{\rm DM}$ as
\begin{eqnarray}
\rho_{\rm DM} 
&=& \rho^{\rm dec}_{3/2} + \rho^{\rm th}_{3/2}
\nonumber \\
&=& f \rho_{\rm pureWDM} + (1-f) \rho_{\rm pureCDM}, 
\end{eqnarray}
where $\rho_{\rm pureWDM}$ and $\rho_{\rm pureCDM}$ are the energy
densities of pure WDM and CDM scenario, respectively.
Considering the adiabatic density fluctuation, we calculate the power
spectrum. As a result, we finally get $f \lesssim 0.4$. In terms of the
density parameter, the constraints indicates $\Omega_{3/2}^{\rm dec}
\lesssim 0.4 \Omega_{\rm DM}$, which gives the upper bound on the
gravitino mass as $m_{3/2} \lesssim 40$ GeV and $m_{3/2} \lesssim 4$ GeV
for the cases of $C_{\rm model} = 1$ and 10, respectively (the light
shaded regions in Fig.\ref{fig:MgvsMbinoV2-4}).

The constraints do not depend highly on the detail of the
observational data.  In fact, in other recent observations, it is
mentioned that the observational error on the power spectrum is around
15\% \cite{Seljak:2004xh,Viel:2005qj}, leading to the constraint as $f
\lesssim 0.2$, which is of the same order of magnitude as the result
above. Therefore, our simple analysis is expected to give the viable
constraint of the structure formation on the WDM+CDM scenario.

\section{Conclusions}

In this paper, we have studied the cosmological implications of the
gravitino LSP scenario with the right-handed sneutrino NLSP in the
framework where neutrino masses are purely Dirac-type. In the case
that MSSM-LSP is Bino-like neutralino, it mainly decays into the
right-handed sneutrino with the lifetime $\tau_{\tilde{B}} \sim
10^2$-$10^3$ seconds in the wide range of the parameter region. Though
the MSSM-LSP is long-lived, no visible particles are produced in the
leading process, thus constraints from the BBN scenario is drastically
relaxed compared to the case without the right-handed sneutrino
NLSP. With the quantitative analysis of the BBN constraints, we have
found the new allowed region, 0.1 GeV $\lesssim m_{3/2} \lesssim$ 40
GeV, when $m_{\tilde{\nu}_R} = $ 100 GeV. In this region, the BBN
constraints give the upper bound on the Bino mass as $m_{\tilde{B}}
\lesssim$ 200-400 GeV, which mainly comes from hadronic four-body
decays. On the other hand, the upper bound on the gravitino mass is
given by the constraints from the structure formation of the
universe. In our scenario, some part of the gravitino is produced by
the decay of right-handed sneutrino at the late universe. As a result,
the gravitino freely streams in the universe and acts as a WDM. The
gravitino is also produced from thermal scattering processes, which
acts as a CDM. Taking the CDM contribution into account, we have
considered the constraints on the WDM+CDM scenario from the
observations of (small scale) structure formation, and finally found
the upper bound on the gravitino mass.

\section*{Acknowledgments}

This work was supported in part by Research Fellowships of the Japan
Society for the Promotion of Science for Young Scientists (K.I.), and
by the Grant-in-Aid for Scientific Research from the Ministry of
Education, Science, Sports, and Culture of Japan, No.\ 19540255
(T.M.).

\end{document}